# Subdiffusion of volcanic earthquakes

Sumiyoshi Abe[1,2,3] and Norikazu Suzuki[4]

[1] Physics Division, Faculty of Information Science and Engineering, Huaqiao University, Xiamen 361021, China

[2]Department of Physical Engineering, Mie University, Mie 514-8507, Japan

[3]Institute of Physics, Kazan Federal University, Kazan 420008, Russia

[4]College of Science and Technology, Nihon University, Chiba 274-8501, Japan

**Abstract** A comparative study is performed on volcanic seismicities at Icelandic volcano, Eyjafjallajökull, and Mt. Etna in Sicily from the viewpoint of complex systems science, and the discovery of remarkable similarities between them is reported. In these seismicities as point processes, the jump probability distributions of earthquakes (i.e., distributions of the distance between the hypocenters of two successive events) are found to obey the exponential law, whereas the waiting-time distributions (i.e., distributions of inter-occurrence time of two successive events) follow the power law. A careful analysis is made about the finite size effects on the waiting-time distributions, and the previously reported results for Mt. Etna (Abe and Suzuki 2015) are reinterpreted accordingly. It is shown that the growth of the seismic region in time is subdiffusive at both volcanoes. The aging phenomenon is commonly observed in the "event-time-averaged" mean-squared displacements of the hypocenters. A comment is also made on (non-)Markovianity of the processes.

## 1. Introduction

Deeper understanding of volcanic seismicity is not only of academic interest but also of obvious importance for mitigation of disaster by eruption, since volcanic earthquakes always occur prior to eruption, although eruption does not necessarily take place after their occurrence.

Traditional geophysical approaches suggest that actually this phenomenon is challenging also for science of complex systems. They involve complexity of the high-level regarding accumulation of stress at faults inside a volcano due to quite a few factors including propagating/inflating/magma-filled dikes, groundwater transport through porous media, and nontrivial geometry of the shape of magma migration (Roman and Cashman 2006; Turcotte 1997; Zobin 2012). Thus, volcanic seismicity results from interplay of diverse complex dynamics on complex architecture at various scales.

It should also be noted that the problem of earthquake-volcano coupling is equally challenging even for non-volcanic earthquakes (Hill et al. 2002).

In a situation where fundamental dynamics governing a system under consideration is largely unknown, the first step toward extracting information about it is to phenomenologically characterize the properties of correlation contained. For this purpose, we have recently studied the feature of diffusion of volcanic earthquakes at Mt. Etna (Abe and Suzuki 2015). There, we have discovered that the volcanic seismicity as a point process exhibits subdiffusion. To understand this behavior, we have analyzed the spatio-temporal properties of the process and have found that the jump probability distribution (i.e., distribution of the distance between the hypocenters of two successive events) obeys the exponential law, whereas the waiting-time distribution (i.e., distribution of inter-occurrence time, or calm time, of two successive events) follows the power law. These results suggest that the spatial aspect of the process is normal in spite of the presence of the complex architecture, and the anomaly originates from the nontrivial nature of the temporal aspect, although there are no physical reasons for separability of these two, in general.

Now, after the work (Abe and Suzuki 2015), we have noticed that the finite size effects may play a significant role in the discussion about diffusion of volcanic earthquakes and, in particular, the value of the exponent in the power-law waiting-time distribution can be sensitive to the effects. To clarify this point, it is desirable to develop a comparative study of seismicities at Mt. Etna and other volcanoes.

The purpose of the present work is to execute such a comparative study on the volcanic seismicities at the volcano, Eyjafjallajökull, in Iceland and Mt. Etna in Sicily. We carefully examine the finite size effects and generalize our previous results (Abe and Suzuki 2015). As will be seen, they share the remarkable common natures in their seismicities as point processes. Both of them are subdiffusive with the values of the anomalous diffusion exponent close to each other. The jump probability distributions are of the exponential type, whereas the waiting-time distributions obey the power law, the exponent of which turns out to be sensitive to the data size. In addition, for both volcanoes, the aging phenomenon is observed for the “event-time-averaged” mean-squared displacements. We also make a comment on existence of the long-term memory effect in the context of violation of the scaling relation to be satisfied by a class of singular Markovian processes.

This paper is organized as follows. In Sec. 2, the subdiffusive nature of volcanic seismicity as a point process is demonstrated. In Sec. 3, the jump probability distributions of the volcanic earthquakes are analyzed. The aging phenomenon is also discussed, there. In Sec. 4, the waiting-time distributions are studied. There, it is shown how the power-law exponent is sensitive to the data size, in general. In Sec. 5, a comment is made on the (non-)Markovian nature of the volcanic seismicity. Sec. 6 is devoted to concluding remarks.

The datasets employed in the present work are available at: (A) http://hraun.vedur.is/ja/viku for Eyjafjallajökull and (B) http://www.ct.ingv.it/ufs/analisti/catalogolist.php for Mt. Etna.

## 2. Subdiffusion in Volcanic Seismicities

First of all, we present in Fig. 1 the patterns of spread of volcanic earthquakes at Eyjafjallajökull and Mt. Etna. There, one recognizes the signs of diffusion. However, the comparison between Fig. 1 (A) and (B) shows that the pattern at Eyjafjallajökull is visually more heterogeneity than the one at Mt. Etna. This might be due to the fact that, in contrast to Mt. Etna, Eyjafjallajökull is not isolated, being with neighboring volcanoes. Accordingly, one would naively imagine that significant differences may exist in the spatio-temporal natures of their seismicities. Quite remarkably, however, they turn out to share similar properties.

We note that Fig. 1 plots the epicenters, but in what follows we will deal with the hypocenters.

As in the work (Abe and Suzuki 2015), we characterize these diffusive patterns in the following way. Consider a sphere with radius $l$ at time $t$ that encloses all earthquakes occurred during the time interval $[0, t]$, where the initial time, 0, is adjusted to the occurrence time of the first event in a point process defined by a sequence extracted from the data under consideration. The collection of such spheres should be concentric with the center being fixed at the hypocenter of the first event in the sequence. In other words, $l$ at $t$ is the largest value among the distances of all events from the first event. This idea is analogous to the concept of *mean maximal excursions* (Tejedor et al. 2010). Then, we describe the diffusion property as follows:

$$l \sim t^{\alpha}, \tag{1}$$

where $\alpha$ is a positive constant termed the diffusion exponent. Here, we are using the notation slightly different from that in our previous paper (Abe and Suzuki 2015). Familiar normal diffusion has $\alpha = 1/2$, whereas $\alpha \neq 1/2$ in anomalous diffusion (Bouchaud and Georges 1990; Metzler et al. 2014): subdiffusion (superdiffusion) if $\alpha < 1/2$ ($\alpha > 1/2$).

In Fig. 2, we present the plots of the law in Eq. (1) for the volcanic seismicities at (A) Eyjafjallajökull and (B) Mt. Etna. The datasets employed here are as follows.

(A-1) During 19:34:21.840 on 2 March 2010 and 08:15:57.078 on 23 July, 2010, 63.503° N – 63.750° N latitude and -19.749° E – -19.024° E longitude.

The total number of events contained is 4000.

(A-2) During 19:34:21.840 on 2 March 2010 and 04:28:6.578 on 16 March, 2014, 63.415° N – 63.750° N latitude and -19.888° E – -18.754° E longitude.

The total number of events contained is 12000.

(B-1) During 22:34:14 on 12 July, 2001 and 01:25:36 on 20 February, 2002, 37.533° N–37.890° N latitude and 14.826° E–15.277° E longitude.

The total number of events contained is 600.

(B-2) During 22:34:14 on 12 July, 2001 and 16:26:35 on 7 Jun, 2010, 37.509° N–37.898° N latitude and 14.706° E–15.298° E longitude.

The total number of events contained is 5000.

As can be seen there, the values of the diffusion exponent are much smaller than 1/2. Therefore, we conclude that the volcanic seismicities exhibit subdiffusion. We note that the growth of $l$ is discontinuous and the step-like behavior appears. $l$ remains constant for a certain duration of time and then abruptly increases. Because of the finiteness of the volcanoes in size, the horizontal length of step becomes unboundedly large in a later stage and then $l$ stops increasing in time, then. Such a stage is irrelevant to diffusion. This point can clearly be seen for Mt. Etna, since it is isolated. However, the situation is much more involved in the case of Eyjafjallajökull, which has the neighborhoods. These issues lead to importance of the finite size effects in the diffusion processes, as will be seen below.

Closing this section, we emphasize that the data intervals mentioned above will always be fixed in the subsequent analysis.

## 3. Spatial Properties

Here, firstly we discuss the jump probability distribution $P_J(\rho)$, where $\rho$ is the three-dimensional distance between two successive earthquakes. In the work (Abe and Suzuki 2003), we have studied this quantity for non-volcanic seismicity and have found that it obeys a statistically exotic law characterized by the distribution decaying faster than the exponential-class ones. However, in the case of volcanic seismicity, it turns out not to be exotic at all.

In Fig. 3, we present the plots of the jump probability distributions for the datasets mentioned in Sec. 2. The result shows that it well obeys the exponential law

$$P_J(\rho) \sim \exp\left(-\rho / \rho_0\right) \tag{2}$$

at both volcanoes. Here, $\rho_0$ is a positive constant having the dimension of length and its values are given in the caption. This quantity may give information about the spatial scale of each volcano. However, we note that its value depends on the data size. In fact, $\rho_0$ in (B-1) is different from that given in our previous work (Abe and Suzuki 2015).

The law in Eq. (2) means that the spatial property of the process is not anomalous and no long jumps are important in contrast e.g. to Lévy flights (Shlesinger et al. 1995). Since long jumps enhance diffusion, their absence is consistent with subdiffusion concluded in the preceding section. However, the final conclusion should not be made until the temporal property of the process is examined.

Secondly, we discuss nonstationarity of the process. For this purpose, we take the series $\{\mathbf{r}_k\}_{k=0,1,2,\ldots}$ with $\mathbf{r}_k$ being the hypocenter of the $k$-th event. The index $k$ plays a role of the internal time referred to as *event time*. Then, we consider the event-time-averaged mean-squared displacement defined by (Abe and Suzuki 2015)

$$\overline{\delta^2(n;a,N)} = \frac{1}{N-n} \sum_{m=a}^{a+N-n-1} \left(\mathbf{r}_{m+n} - \mathbf{r}_m\right)^2 , \tag{3}$$

where $a$, $n$ and $N$ are referred to as the *aging event time*, *lag event time* and *measurement event time*, respectively, provided that $N-n$ should be much larger than unity. [The upper limit of the summation on the right-hand side of Eq. (6) in the paper (Abe and Suzuki 2015) should read as the one in Eq. (3) above, but this change is negligibly small for the result given there.] $\overline{\delta^2}$ can be regarded as a discrete event-time version of the time-averaged mean-squared displacement studied in the recent works (Metzler et al. 2014; Schulz et al. 2014).

In Fig. 4, we present the plots of the quantity in Eq. (3) for some values of the aging event time. The data intervals considered here are (A-2) and (B-2) mentioned in the preceding section. The aging phenomenon is clearly observed at both Eyjafjallajökull and Mt. Etna: That is, monotonic increase of $\overline{\delta^2}$ with respect to the aging event time. This fact, originally found for Mt. Etna (Abe and Suzuki 2015), implies that the sequence of the hypocenters belongs to a specific class of *nonstationary* point processes.

## 4. Temporal Properties

Since the jump probability distribution is not anomalous, the subdiffusive nature discussed in Sec. 2 is supposed to be concerned with long waiting time $\tau$ between two successive events that suppresses diffusion. In this section, we show that this is indeed the case.

In Fig. 5, we present the plots of the waiting-time distributions. Once again, the data intervals employed here are the same as the ones in Sec. 2. As expected from the above consideration, the distributions, in fact, decay as the power law

$$P_W(\tau) \sim \tau^{-1-\mu}. \qquad (4)$$

Here, the notation different from that in the paper (Abe and Suzuki 2015) is used for the exponent. In this respect, it may be worth mentioning that the power-law waiting-time distribution is observed also for non-volcanic earthquakes (Abe and Suzuki 2005).

It is of importance to note that the value of the exponent is different for different size of data interval. In Fig. 6, we show how $\mu$ depends on the data size. There, we see a remarkable similarity between Eyjafjallajökull and Mt. Etna, apart from the fact that the volcanic seismicity at Eyjafjallajökull is much more active than that at Mt. Etna in the data intervals considered. We wish to point out that both of the values of $\mu$ at Eyjafjallajökull and Mt. Etna seem to converge to a common value $\mu \cong -0.14$, as the size of the interval increases, suggesting the existence of *universality* in a certain sense.

We wish to emphasize that, in the above, “size” is the length of the conventional time interval of the data and is not of the event time. According to our analysis, as long as the number of events is employed, the data collapse as in Fig. 6 cannot be established. As mentioned above, the volcanic seismicity at Eyjafjallajökull is much more active than that at Mt. Etna. This implies that the event time as the internal time reveals the difference between these two seismicities. Further study is needed for deeper understanding of this issue regarding the concept of time in complex systems.

## 5. A Comment on (Non-)Markovianity

Another temporal property of interest is the decay rate of the number of events. Let $N(t)$ be the number of events occurred in the time interval $[t_0, t]$, where $t_0$ is a fixed time. We are particularly interested in data intervals, in which the rate decays as the power law: $dN(t)/dt \sim A/t^p$, where $p$ and $A$ are positive constants. This may remind one of the Omori-Utsu law (Omori 1894; Utsu 1961) for the temporal pattern of aftershocks following a main shock. We note, however, that in the present case the initial event in the data interval under consideration does not necessarily correspond to the one with a large value of magnitude. Now, it is convenient to employ the integrated form of the law:

$$N(t) - N(t_0) \sim \begin{cases} A\left(t^{1-p} - t_0^{1-p}\right)/(1-p) & (p \neq 1) \\ A\ln(t/t_0) & (p = 1) \end{cases} . \qquad (5)$$

In Fig. 7, we show how the frequency of occurrence of events in the dataset (B-1) in Sec. 2 varies in time. There, a noticeable trend can be recognized in the subinterval indicated by the left-right arrow. (On the other hand, no such trends could be found for Eyjafjallajökull. Accordingly, here we only analyze the data taken from Mt. Etna.) In Fig. 8, we present the plots of both Eq. (5) and the waiting-time distribution for the subinterval in Fig. 7. (Recall that the value of the exponent of the power-law waiting-time distribution depends on the data size.) It is mathematically known (Barndorff-Nielsen et al. 2000) that, for a class of singular Markovian processes with both $p$ and $\mu$ being in the range $(0,1)$, holds the following scaling relation:

$$p + \mu = 1 . \tag{6}$$

In other words, violation of this relation implies that the system has long-term memory. Now, from Fig. 8, we approximately estimate $p + \mu \cong 1.23$. Thus, the process in the data interval under consideration seems to be non-Markovian. This point is analogous to that in non-volcanic seismicity. We have previously studied a similar issue for earthquake aftershocks (Abe and Suzuki 2009; Abe and Suzuki 2012). There, we have reported significant violation (~50%) of Eq. (6), leading to the fact that any model assuming Markovianity of aftershock sequence should be reconsidered. However, the violation of the scaling relation here is not so significant.

As additional information, we also point out that Eq. (6) has also been discussed in other contexts including laser cooling of atoms (Bardou et al. 2002) and acoustic emissions from plunged granular materials to examine (non-)Markovianity (Tsuji and Katsuragi 2015).

As mentioned above, we could not apply the scaling method to the volcanic seismicity of the data intervals considered here for Eyjafjallajökull, and therefore other approaches are needed for examining (non-)Markovianity for its seismicity. The best we can say at present is that volcanic seismicity seems to possess ~~has~~ elements of non-Markovianity, in general, and this point is in consistent with the complex natures of the phenomenon that we have observed in the present work.

## 6. Concluding Remarks

We have performed a comparative study of diffusion of volcanic earthquakes at Eyjafjallajökull and Mt. Etna. We have found that at both of them the phenomenon is subdiffusive characterized by the values of the anomalous diffusion exponent close to each other. Then, we have analyzed the spatio-temporal properties of the volcanic seismicities as point processes. We have shown that the jump probability distributions for these volcanoes obey the exponential law, whereas the waiting-time distributions do the power law. We have examined how the exponent of the power-law waiting-time distribution is sensitive to the data size. These show that the seismicities at these volcanoes share the remarkable common features, indicating universalities of the findings presented here. Furthermore, we have also analyzed the occurrence rate of volcanic earthquakes in time to examine if the volcanic seismicity is (non-)Markovian.

Previously, we have discussed (Abe and Suzuki 2015) that all of four celebrated theoretical approaches to anomalous diffusion, i.e., fractional kinetics of continuous-time random walks, fractional Brownian motion, fractal random walks, and nonlinear kinetics, do not seem to offer a deciding explanation of subdiffusion of volcanic earthquakes. The present work, however, shows that the value of the exponent of the power-law waiting-time distribution is sensitive to the data size, in general, implying that further studies are needed. We would also like to stress that any new theoretical development in anomalous diffusion, e.g., a recent work (Boon and Lutsko 2017), may contribute to understanding the physics of volcanic seismicity.

## Acknowledgments

The works of SA and NS have been supported in part by the Grant-in-Aid for Scientific Research from the Japan Society for the Promotion of Science under the contracts (No. 26400391 and No. 16K05484). SA has also been supported in part by the High-End Foreign Expert Program of China and by the Program of Competitive Growth of Kazan Federal University by the Ministry of Education and Science of the Russian Federation.

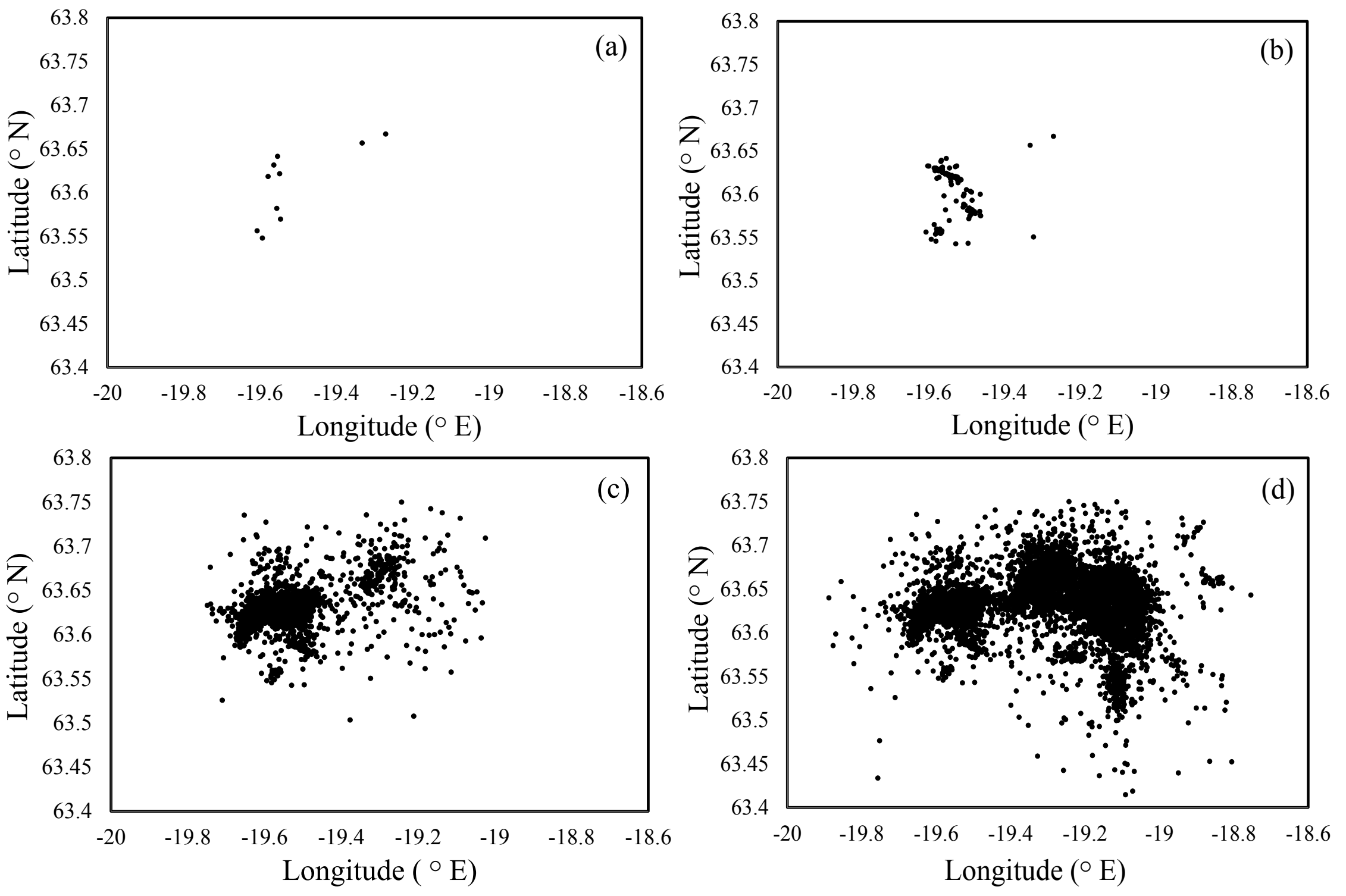

(A)

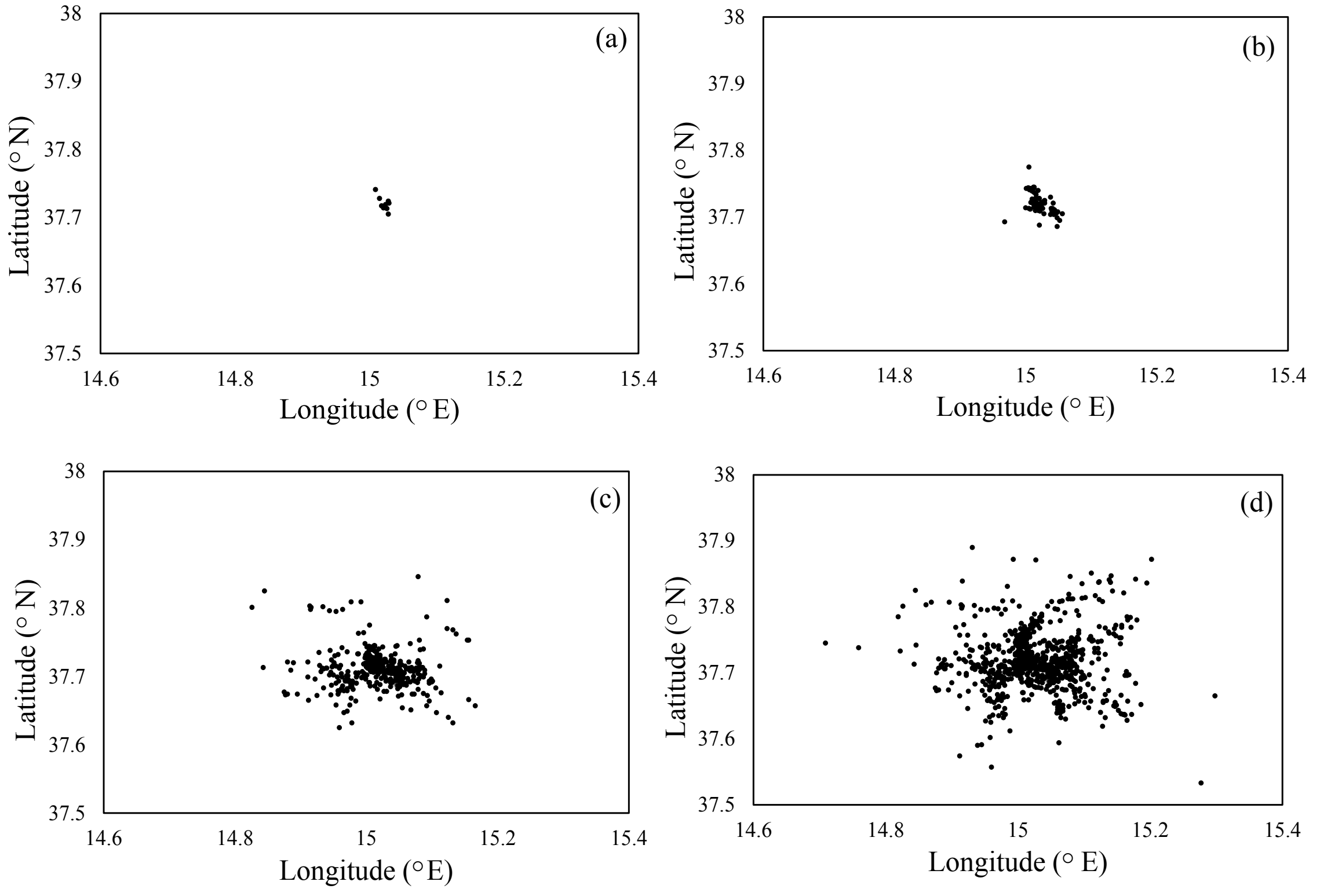

(B)

**Figure 1** The plots of the epicenters of the volcanic earthquakes.
(A) Eyjafjallajökull with the initial event at 19:34:21.84 on 2 March 2010 [the first (a) 10, (b) 100, (c) 4000, (d) 12000 events] and
(B) Mt. Etna with the initial event at 22:34:14 on 12 July, 2001 [the first (a) 10, (b) 100, (c) 500, (d) 1000 events].

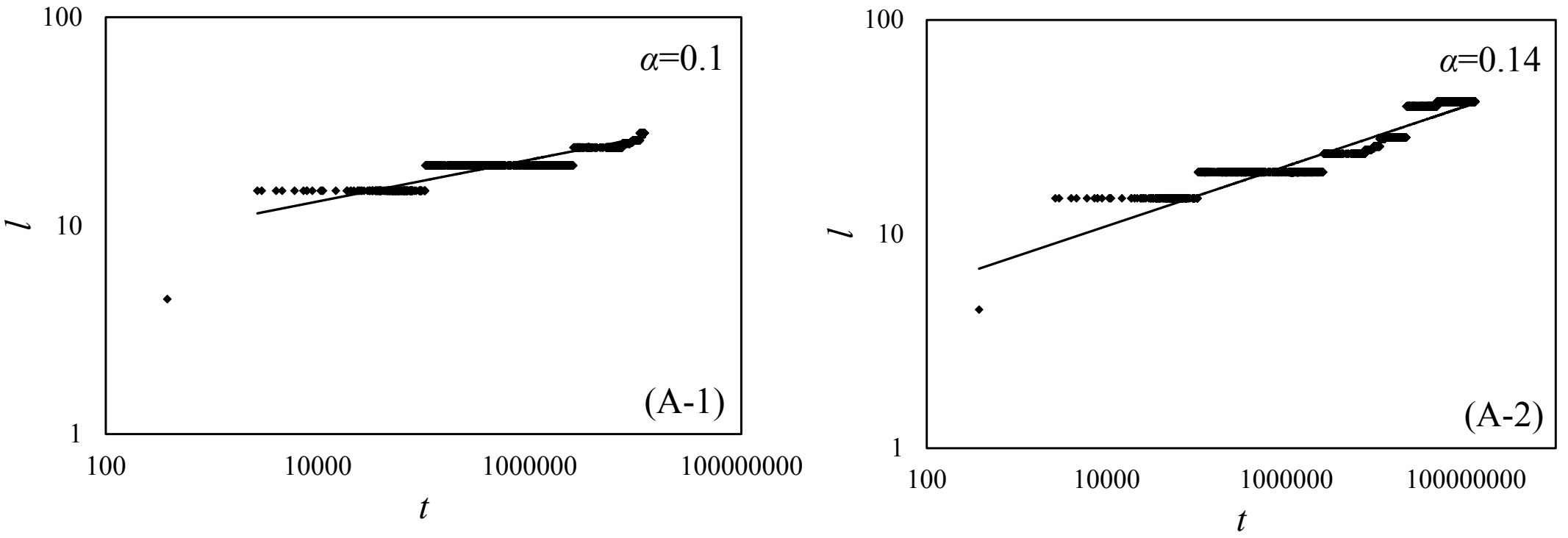

(A)

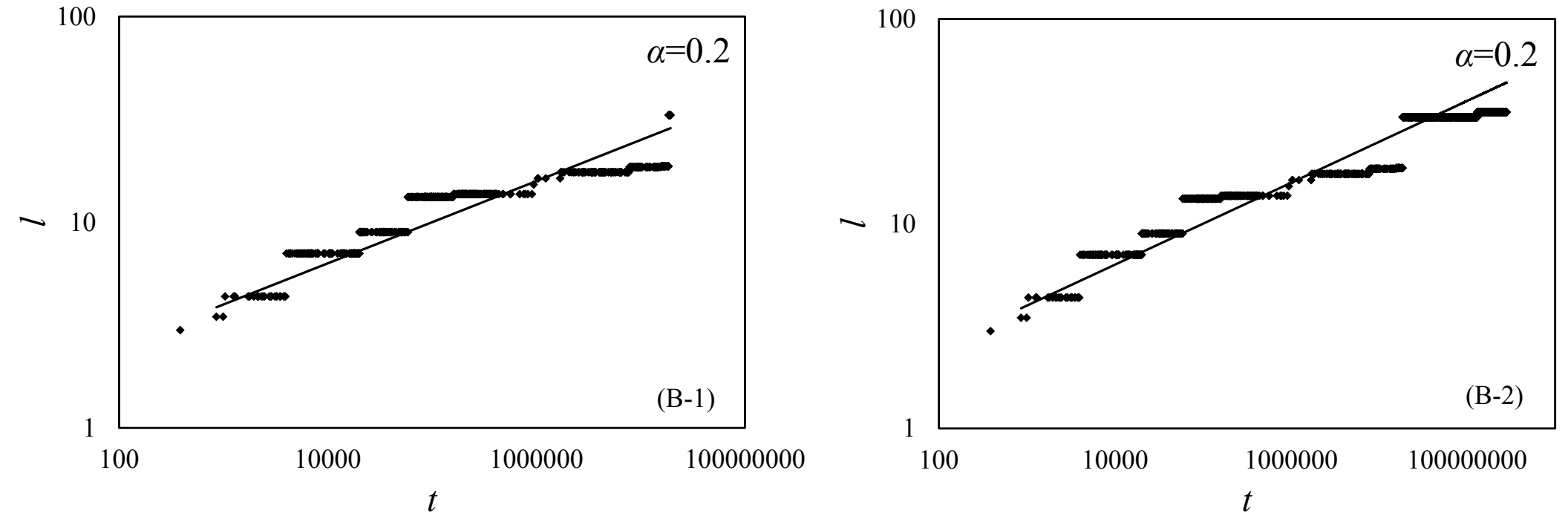

(B)

**Figure 2** The log-log plots of the radius $l$ [km] with respect to (conventional) time $t$ [sec]. The straight lines describe the law in Eq. (1).

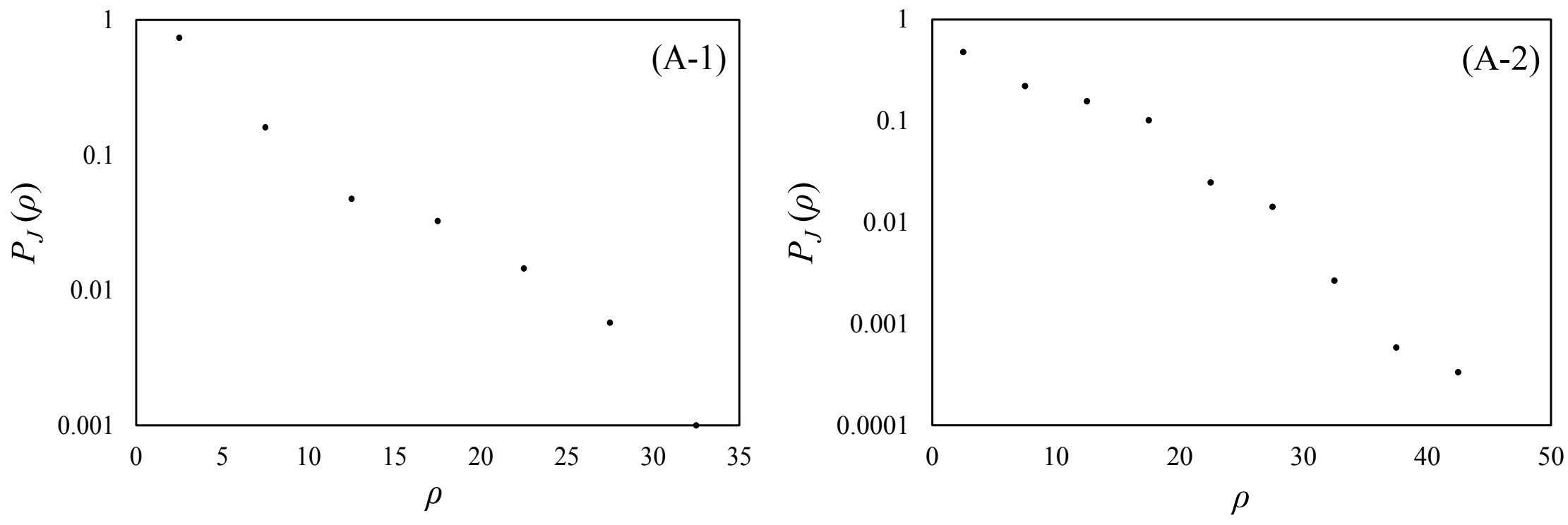

(A)

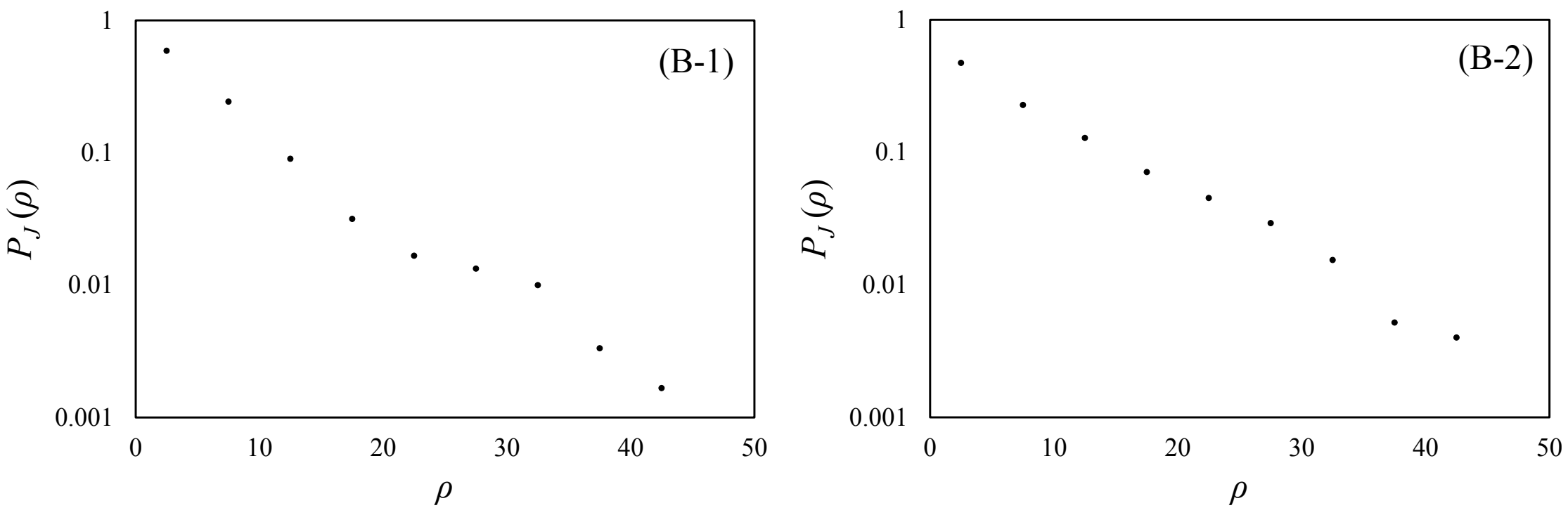

(B)

**Figure 3** The semi-log plots of the normalized jump probability distributions $P_J(\rho)$ [1/km] with respect to $\rho$ [km]. The values of $\rho_0$ in Eq. (2) are: (A-1) 5.0 km, (A-2) 5.0 km, (B-1) 7.5 km, (B-2) 8.5 km. The histograms are made with the bin size 5.0 km.

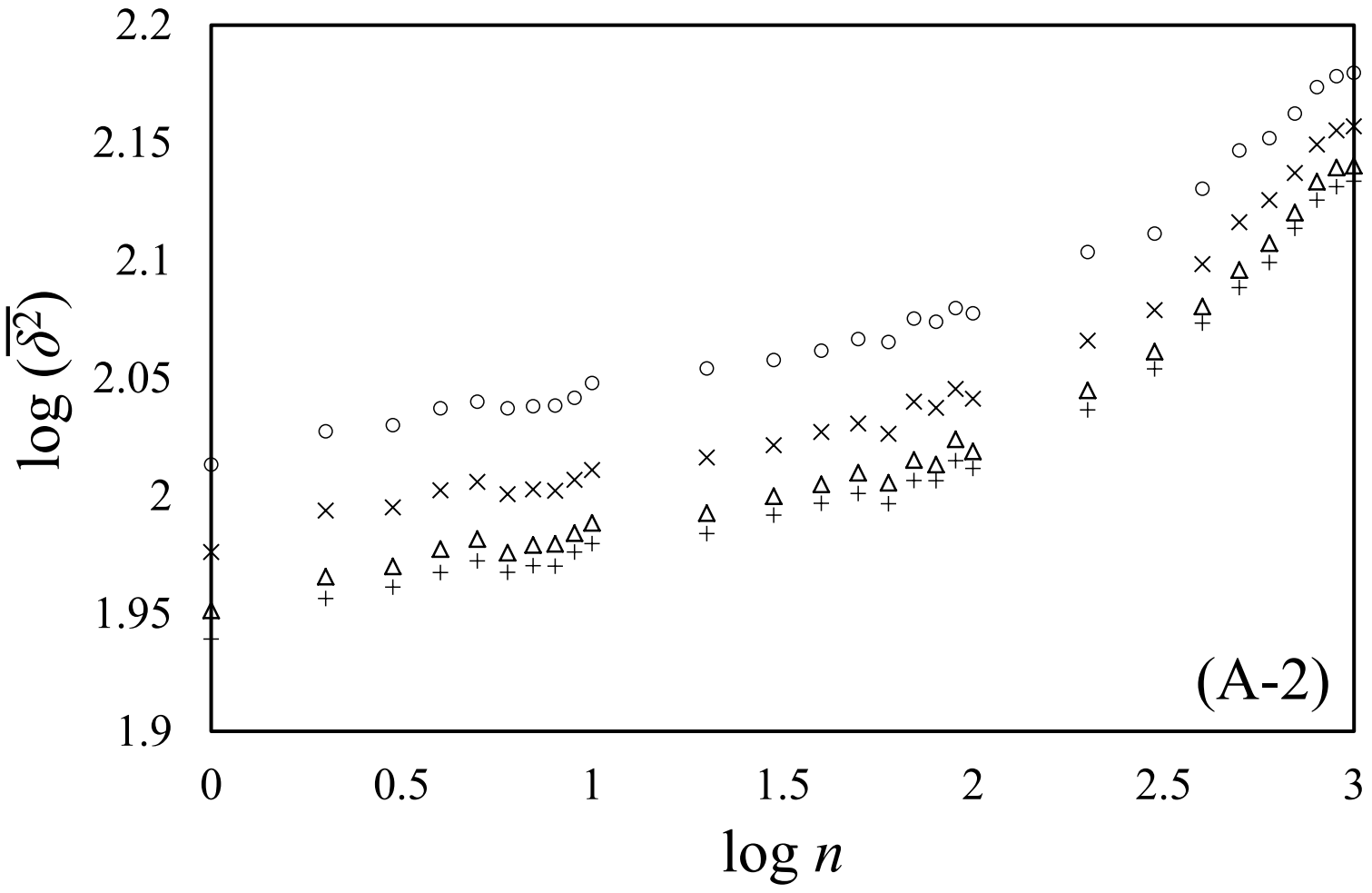

(A)

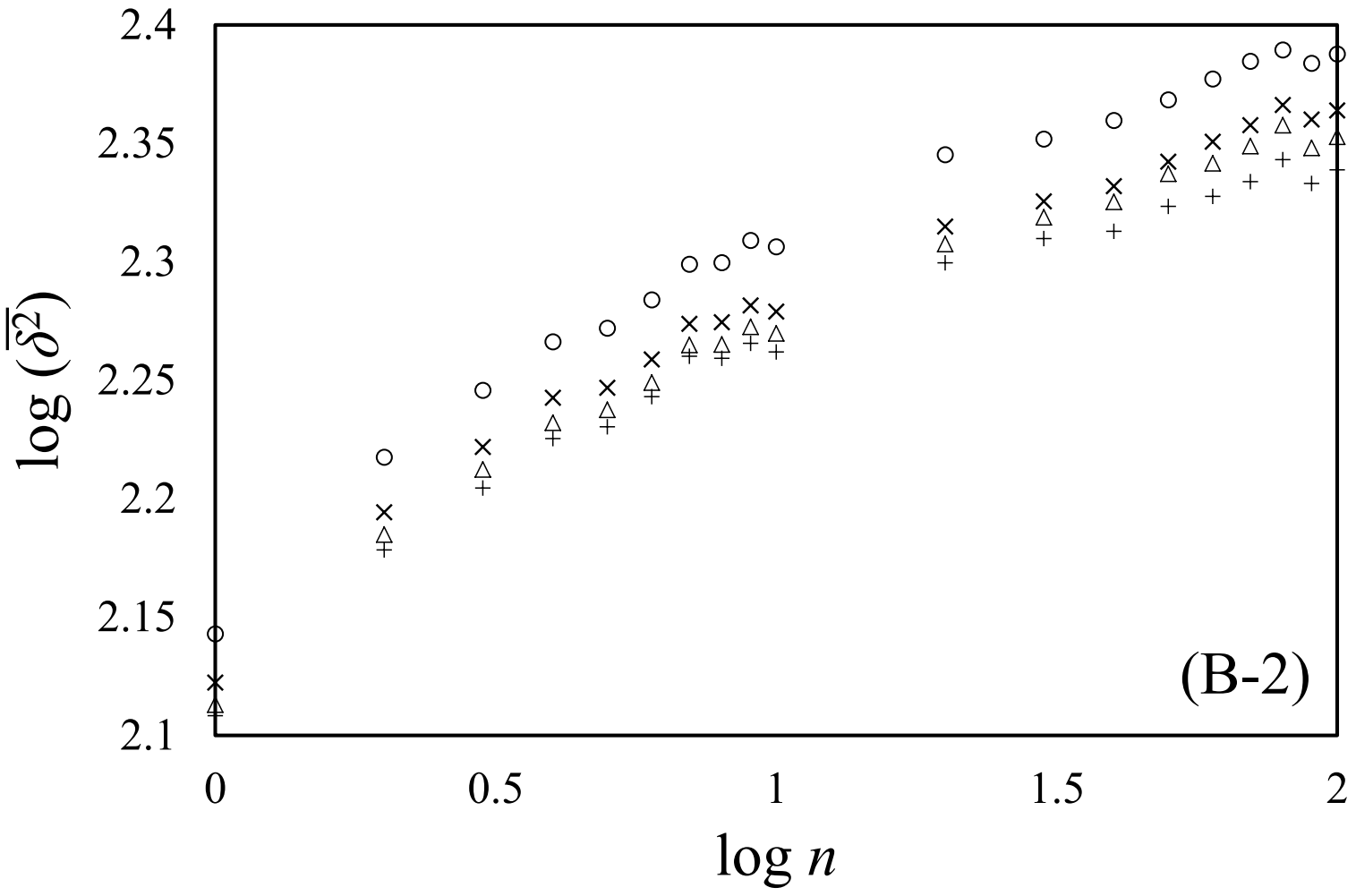

(B)

**Figure 4** The log-log plots of the event-time-averaged mean-square displacement. The values of the aging $a$ and the measurement event time $N$ are: (A) +($a=0$), △($a=200$), ×($a=500$), ∘($a=1000$), $N=11000$ , and (B) +($a=0$), △($a=50$), ×($a=100$), ∘($a=300$), $N=4700$ . All quantities are dimensionless.

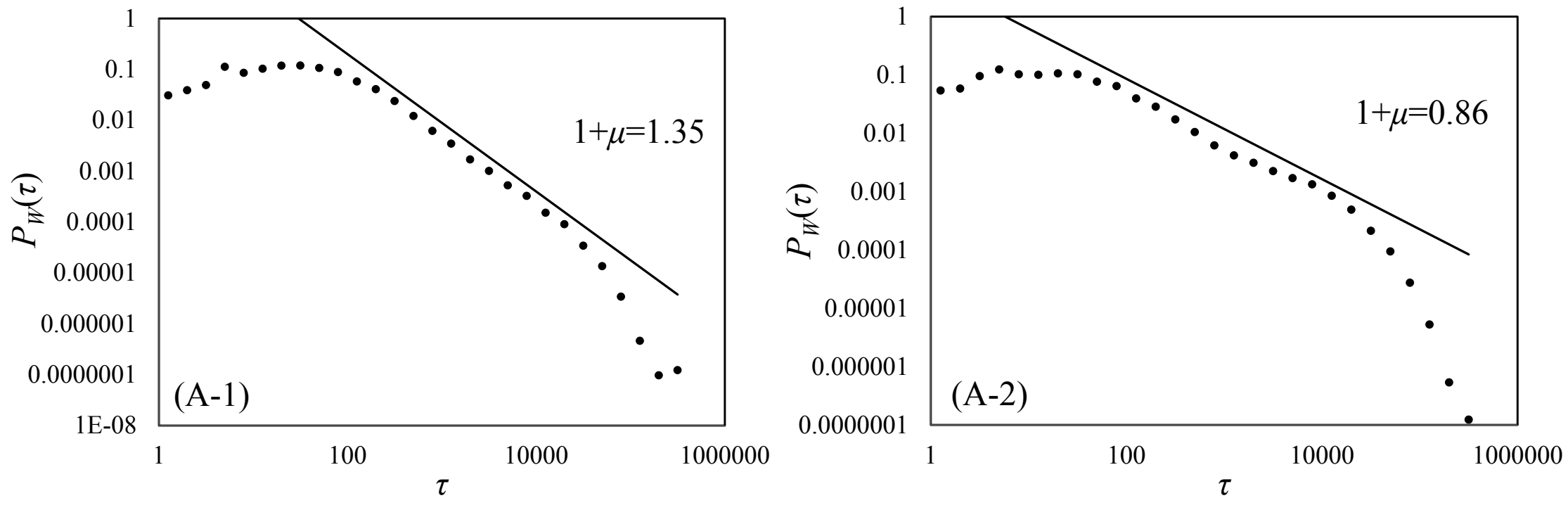

(A)

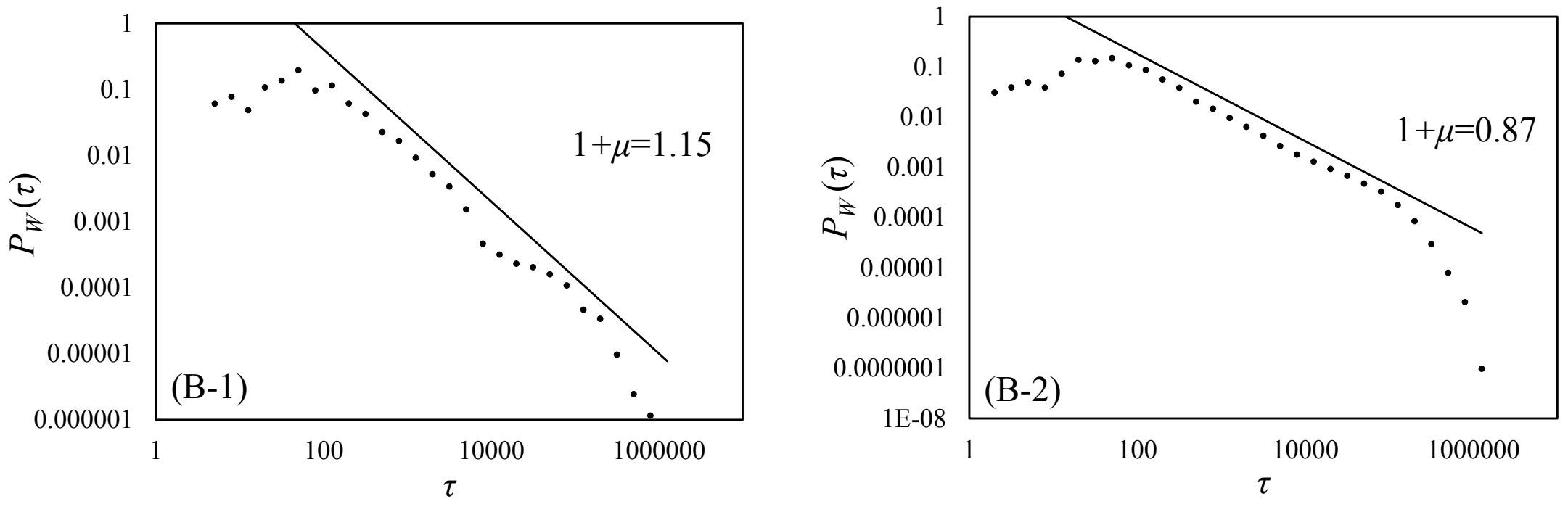

(B)

**Figure 5** The log-log plots of the normalized waiting-time distribution $P_W(\tau)$ [1/sec] with respect to the waiting time $\tau$ [sec]. The bin size for making the histograms is fixed in such a way that each decade contains five points.

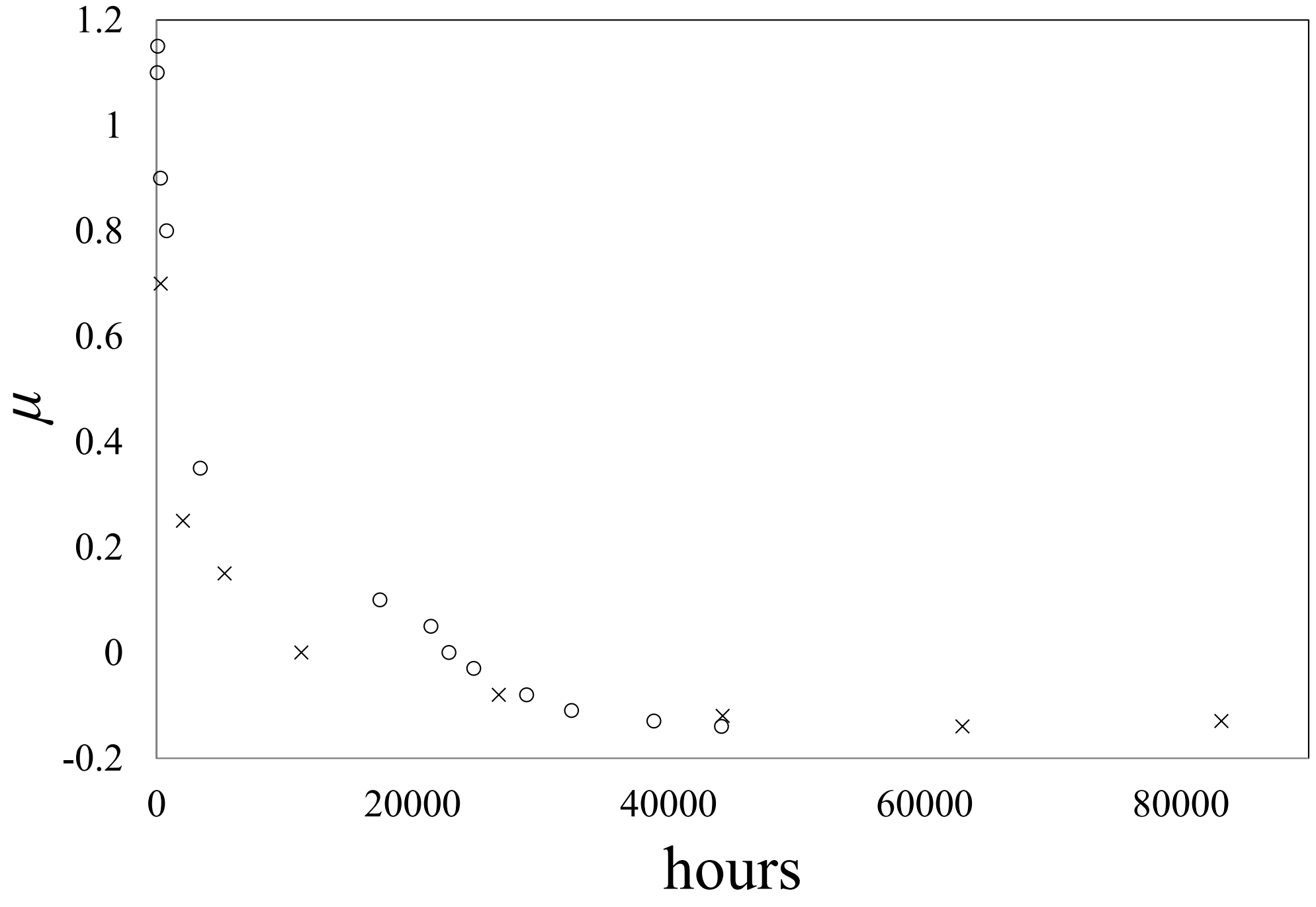

**Figure 6** Dependence of the exponent $\mu$ of the waiting-time distributions on the size of data intervals measured in the unit of 1 hour from the initial events given in Sec. 2. ∘ (Eyjafjallajökull) and × (Mt. Etna). Possible error bars are not indicated, since a primary concern here is to show the gross behavior of $\mu$ with respect to the size of the data intervals.

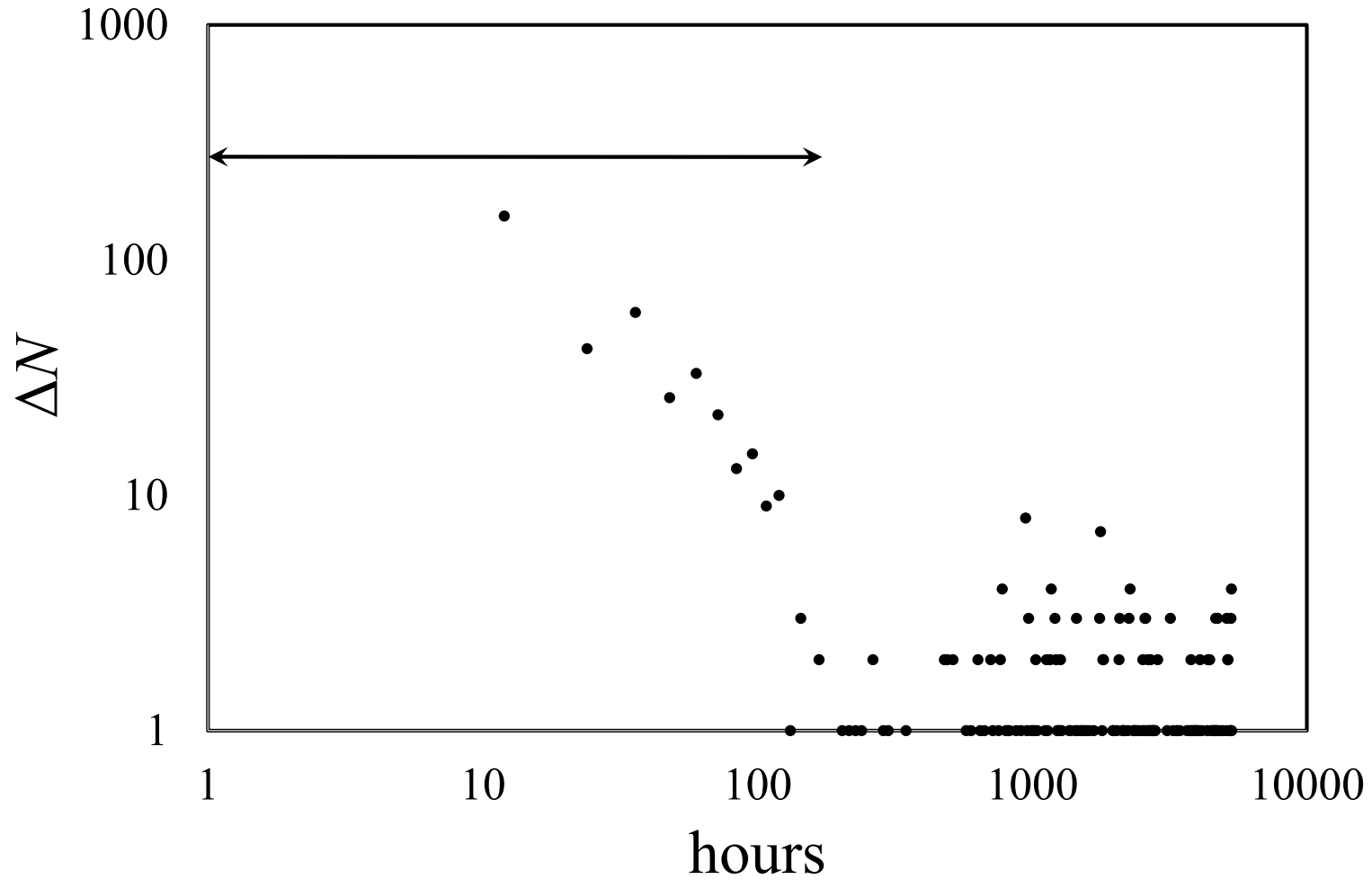

**Figure 7** The log-log plot of the number of events occurred per 12 hours, $\Delta N$, at Mt. Etna with respect to the duration of time measured in the unit of 1 hour from the initial events given in Sec. 2. The subinterval indicated by the left-right arrow is between 22:34:14 on 12 July, 2001 and 12:17:05 on 19 July, 2001. The total number of events contained is 390.

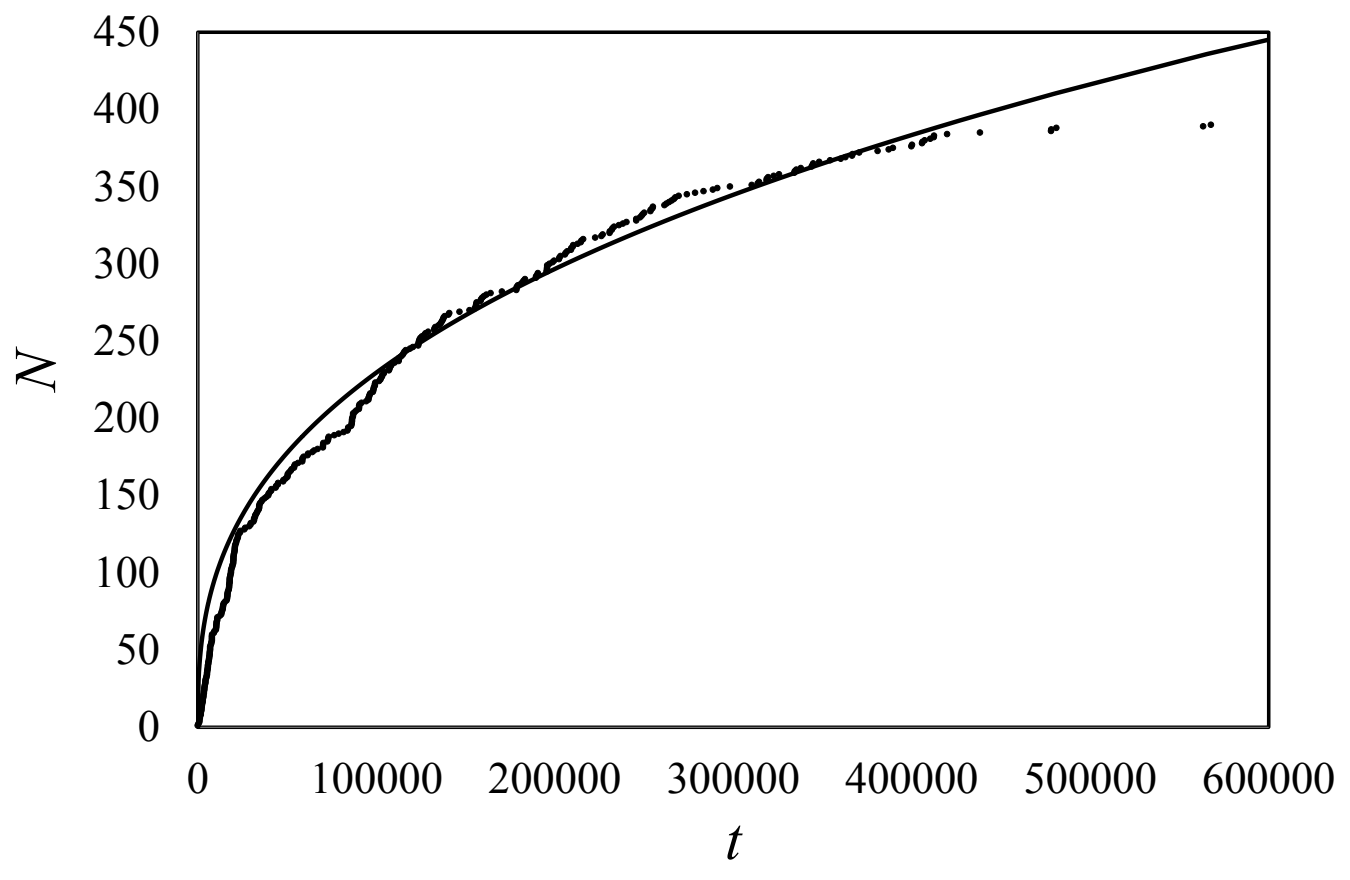

(B-i)

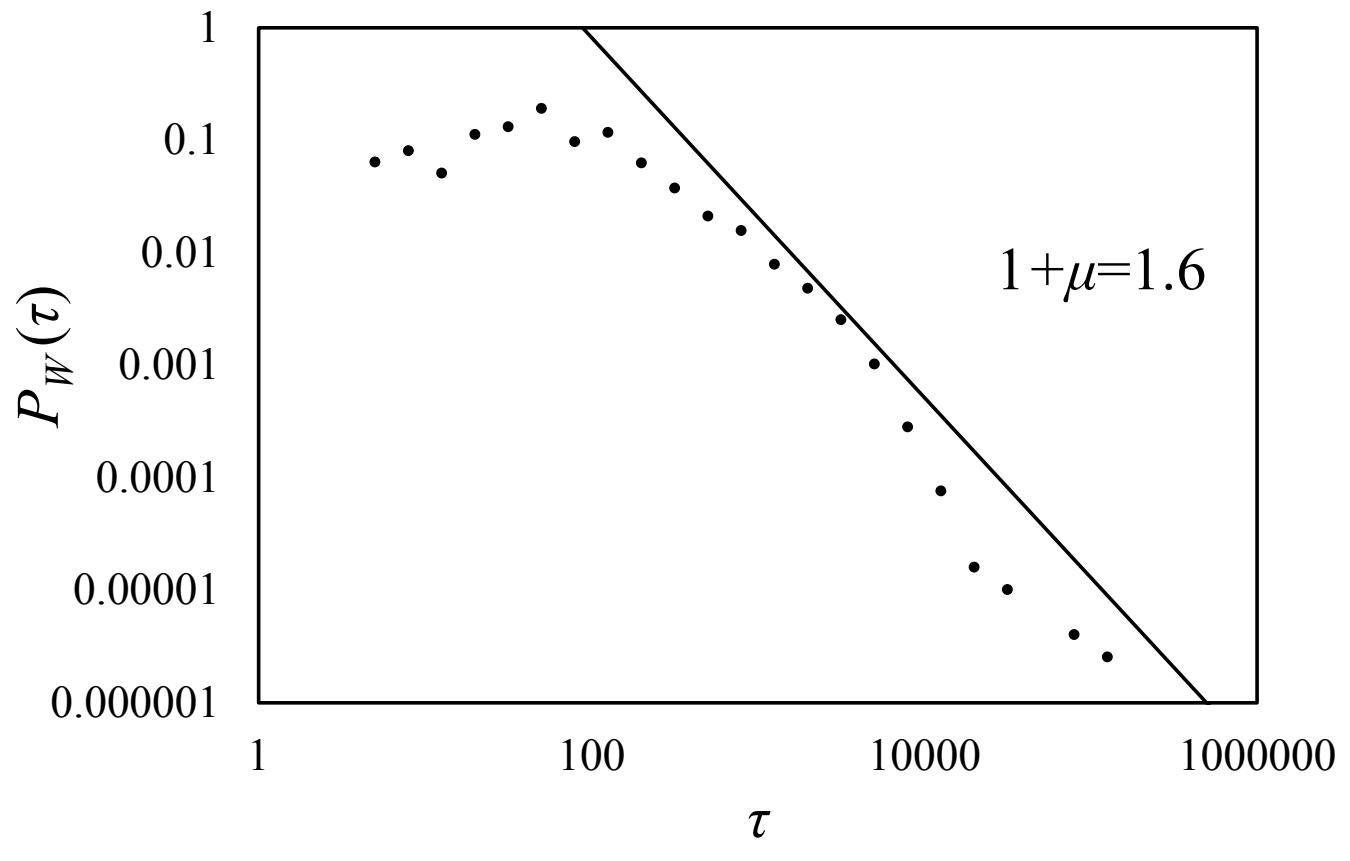

(B-ii)

**Figure 8** The log-log plots of (B-i) the cumulative number of events, $N$, in Eq. (5) with $t_0 \equiv 0$ being adjusted to the occurrence time of the first event contained in the datasets in Sec. 2 and (B-ii) the corresponding waiting-time distribution, $P_W(\tau)$ [1/sec] with respect to the waiting time $\tau$ [sec] in the selected subinterval in Fig. 7. The solid curve in (B-i) describes the law in Eq. (5) with $A \cong 1.20$ $\sec^{p-1}$ and $p \cong 0.63$. The bin size for making the histogram in (B-ii) is fixed in such a way that each decade contains five points.